\documentstyle[12pt]{article}

\begin{document}
\def\beq{\begin{equation}}  
\def\eeq{\end{equation}}
\def\beqa{\begin{eqnarray}}
\def\eeqa{\end{eqnarray}}
\def\noin{\noindent}
\def\grad{{\bf \nabla}}
\def\bv{{\bf v}} 
\def\tb{\tilde{b}}
\def\bB{{\bf B}}
\def\bJ{{\bf J}}
\def\bE{{\bf E}}
\def\bK{{\bf K}}
\def\bx{{\bf x}}
\def\pa{\partial}
\def\pat{{\partial_t}}
\def\eps{\epsilon_{\alpha\beta}}
\def\rta{\rightarrow}
\def\lra{\leftrightarrow}
\def\tm{\tilde{m}}
\def\tn{\tilde{n}}
\def\tbv{\tilde{\bv}}
\def\tbK{\tilde{\bK}}
\def\tP{\tilde{P}}
\def\trho{\tilde{\rho}}
\def\hm{\hat{m}}
\def\hn{\hat{n}}
\def\hbv{\hat{\bv}}
\def\hbK{\hat{\bK}}
\def\hP{\hat{P}}
\def\hrho{\hat{\rho}}
\def\cE{{\bf {\cal E}}}   
\def\cB{{\bf {\cal B}}}   
\def\bbJ{\bar{\bJ}}
\def\brho{\bar{\rho}}
\def\bsig{\bar{\sigma}}
\def\ccB{{\cal B}}       
\def\ccE{{\cal E}}        
\def\ccA{{\cal A}}        
\def\hcA{\hat{\ccA}}
\def\bA{\bf A}
\titlepage
\begin{flushright} QMW-PH-96-10
\end{flushright}
\vspace{3ex}
\begin{center} \bf
{\bf STRINGS AND DYONIC PLASMAS}\\
\rm
\vskip 0.5cm

Omduth Coceal$^{a}$\footnote{e-mail: o.coceal@qmw.ac.uk}
Wafic A. Sabra$^{b}$\footnote{e-mail: 
uhap012@vax.rhbnc.ac.uk} and
Steven Thomas$^{a}$\footnote{e-mail: s.thomas@qmw.ac.uk}\\
\vspace{2ex}
$^{a}${\it Department of Physics\\
Queen Mary and Westfield College\\
Mile End Road\\
London E1 4NS\\
U.K.}\\
\vspace{4ex}
$^{b}${\it Department of Physics\\
Royal Holloway and Bedford New College\\
Egham\\
Surrey TW20 0EX\\
U.K.}\\
\vspace{4ex}
ABSTRACT
\end{center}
\noindent
Recently Olesen has shown the  existence of dual string solutions to the equations of 
ideal Magnetohydrodynamics that describe the long wavelength properties 
of electrically charged plasmas. Here, we extend these solutions to include the case of 
plasmas consisting of point like dyons, which carry both electric and magnetic charge. 
Such strings are dyonic in that they consist of both magnetic and electric 
flux. We contrast some physical features of dyonic plasmas with those of the purely 
electric or magnetic type, particularly in relation to the validity of the ideal 
approximation.
\newpage
Olesen \cite{olesen} has discussed the intriguing possibility
that dual strings could provide solutions to the theory of 3-dimensional 
magnetohydrodynamics. Such strings are associated with 
either purely magnetic or electric flux tubes (the latter being solutions 
of the dual version of MHD). The magnetic flux tubes can arise as
an extension of the 2+1-dimensional Nielsen-Olesen vortex \cite{nov} to 3+1 dimensions. 
Purely magnetic strings are known to form 
in the early universe as topological defects as a consequence of a phase transition 
(see \cite{hindmarsh} for a recent review).
Whilst these strings are approximate solutions of the usual Nambu-Goto equations,
the MHD approach offers an alternative description which might prove useful in 
some circumstances.
  
From an MHD point of view, the existence of such (non-decaying) solutions depends, 
to varying degree, on a number of assumptions concerning the dynamics 
of the string or finite magnetic flux tube as it moves in the plasma. Perhaps 
most notable are the conditions that the velocity of any point on the string is that 
of the ambient fluid in which it moves, and the perpendicularity between the  
velocity $\bv$ and the magnetic field $\bB$. Moreover the hydrodynamical 
system is assumed to be ideal \cite{nicolson}, otherwise the problem 
is complicated by the fact that strings can decay.  

In this letter we want to generalize these string solutions to the case
of a 3-d plasma consisting of dyons, i.e idealized point particles
carrying both electric and magnetic charges. This naturally leads to the  
interpretation of such strings as dyonic flux tubes. An obvious question 
is what field theory models can give rise to such string-like solutions in
4 dimensions, which would then be a dyonic analogue of the extended
Nielsen-Olesen vortex discussed above. Such a generalization is not obvious since
in principle the field theory would include both electric and magnetic sources, 
which is well known to be problematic in general. However, we shall see 
that in the hydrodynamical picture of dyonic plasmas, the electric and magnetic 
currents are not linearly independent, but are proportional to each other. Ultimately, it 
is this relation that explains why the hydrodynamical models of electric and dyonic plasmas
are closely connected [10]. If this feature persists in any potential field theory 
description, then this could provide a way to overcome the well known difficulty of 
introducing  electromagnetic potentials in the presence of both electric and magnetic 
sources. A straightforward generalization of the abelian Higgs model and the Nielsen-Olesen
vortex might then be possible. Whatever form the field theory description of dyonic flux 
tubes takes, it is natural to expect the 
resulting theory to be invariant under electric-magnetic duality symmetry. 
This might give us a clue as to 
another possible origin of dyonic strings, at least if we accept the introduction of 
supersymmetry, because there has been much recent progress and 
mounting evidence that electric-magnetic duality (in the guise of $S$-duality)
is a symmetry for example of $N=4$, $d=4$ heterotic superstring theories \cite{sdual}. 
In passing, we should comment that a family of 
\lq\lq dyonic" string solutions has recently been constructed \cite{schwarz} as 
$d=10$ solitons of the  type IIb superstring theory.
Although these have no obvious connection to the ones we shall discuss in this letter, (since 
in the type IIb theory the 
electric and magnetic charges correspond to different gauge groups), they at least provide 
an anology. 
We shall speculate more on these issues in the conclusion.

In contrast to the above discussions, our original motivation in studying dyonic plasmas in
3-dimensions was to obtain generalized hydrodynamical 2-d systems by dimensional
reduction \cite{ct1,ost2}. Two-dimensional plasmas involving electric, magnetic 
and dyonic charges could then provide 
a further testing ground for the ideas behind the theory of 
conformal turbulence \cite{polyakov}. In particular recent  
numerical studies \cite{benzi} concerning the validity
of conformal turbulence in 2-d fluid flow, could be applied to such systems.
Of course, one would expect that theories of generalized MHD will have 
much  broader physical applications than these.

We begin by considering the effective (i.e long wavelength) properties of
a plasma of dyonic particles, carrying equal and opposite electric and magnetic charges 
$\pm(e,g)$ which was studied in \cite{ost1}. It is important to note that we are 
not considering a plasma of degenerate dyons and anti-dyons having the same mass. Rather 
we imagine a system of dyons with charges $(e,g)$ and mass $m_d$ together with 
their \lq\lq ionic" counterparts carrying charges $(-e,-g)$ and a different mass $m_{di}$. 
This is the direct analogue of usual electric plasmas consisting of charge $e$ electrons of 
mass $m_e$ and $-e$ ions of mass $m_d$. Moreover an important approximation which is 
used  in deriving the long wavelength 
properties of the latter is that  $m_e << m_i$. This hierarchy of mass scales essentially 
allows a 1-fluid description of the plasma and hence makes contact with hydrodynamics.
Such an approximation is clearly valid in realistic 
plasmas.  In our treatment of dyonic plasmas we shall also assume that 
$m_d << m_{di}$ in order to yield a 1-fluid model.

The equations for dyonic plasmas of this kind \cite{ost1}
consist of Maxwell's equations in the presence of both magnetic and electric charges 
and currents
\beqa\label{eq:1}
\grad\cdot\bE& =& \rho_{c},\qquad\qquad \grad\cdot\bB=\rho_{m},\quad\cr
\cr
\grad\times\bB&=&\frac{4\pi}{c}\bJ_{e}+\frac{1}{c}\pat{\bE},\quad  
\grad\times\bE =-\frac{4\pi}{c}\bJ_{m}-\frac{1}{c}\pat{\bB}, 
\eeqa 
where $\rho_c$ and $\rho_m$ are the electric and magnetic charge 
densities, and $\bJ_e,$ $\bJ_m$ the respective electric and magnetic currents.
In addition, one has fluid equations with a suitably generalized Navier-Stokes equation 
\beqa\label{eq:2}
{\dot{\rho}}_{M}& +&\grad\cdot({\rho}_{M}{\bv})  
= 0,\quad
{\dot{\rho}}_{c}+\grad\cdot\bJ_{e}=0,\quad
{\dot{\rho}}_{m}+\grad\cdot\bJ_{m}=0,\cr
\cr
\rho_{M}(\pat{\bv}& +& \bv\cdot\grad \bv)= -\grad P + 
\rho_{c}\bE + \rho_{m}\bB + \frac{1}{c}\bJ_{e}\times\bB -\frac{1}{c} 
\bJ_{m}\times\bE+\sum_{s= d, di}{\bf K}_{s}(x) 
\eeqa
$P$ being the total pressure and $\rho_M$ the 
mass density. The quantities ${\bf K}_s$ represent the change in momentum
of dyonic species $s$ at the point $x$ due to collisions, and 
in general they depend on the velocities of each species.  However, applying the 
reasoning outlined in \cite{nicolson} to the present case, it is expected that 
during scattering
${\bf K}_d =-{\bf K}_{di}$ to a good approximation so that the collision terms vanish 
in eq.(\ref{eq:2}). Finally, in addition to eqs (\ref{eq:1})
and (\ref{eq:2}), one can derive generalized Ohm's laws for the currents $\bJ_e $ and 
$\bJ_m$ respectively 
\beqa\label{eq:3}
\frac{m_{d}m_{di}}{\rho_{M}e^{2}}{\pat{\bJ}}_{e} 
&=&\frac{m_{d}}{2\rho_{M}e}\grad P
+(\bE+\frac{1}{c}\bv\times\bB)+ 
\frac{g}{e}(\bB-\frac{1}{c}\bv\times\bE)\cr
\cr
 & &- \frac{m_{d}}{\rho_{M}ec}\bJ_{e}\times\bB
+ \frac{m_{d}}{\rho_{M}ec}\bJ_{m}\times\bE - 
\frac{1}{\sigma_{e}}\bJ_{e},\cr
\cr
\frac{m_{d}m_{di}}{\rho_{M}g^{2}}{\pat {\bJ}}_{m} 
 &=&  \frac{m_{d}}{2\rho_{M}g}\grad P
+ (\bB-\frac{1}{c}\bv\times\bE) +\frac{e}{g} 
(\bE+\frac{1}{c}\bv\times\bB) \cr
\cr
& +& \frac{m_{d}}{\rho_{M}gc}\bJ_{m}\times\bE
-\frac{m_{d}}{\rho_{M}gc}\bJ_{e}\times\bB+ 
\frac{1}{\sigma_{g}}\bJ_{m}
,\eeqa
where $\sigma_e$ and $\sigma_g$ are the electric and magnetic
conductivities respectively. 
Eqs (\ref{eq:1})-(\ref{eq:3}) constitute the hydrodynamical equations of 
a dyonic plasma which we refer to as dyonic hydrodynamics (DHD) and 
are invariant under the electric-magnetic duality symmetries
\beqa\label{eq:4}
\bE &\rta &\bB,\quad \bB\rta -\bE\quad
\bJ_{e}\rta\bJ_{m},\quad\bJ_{m}\rta -\bJ_{e}\quad
\rho_{c}\rta\rho_{m},\quad\rho_{m}\rta -\rho_{c}\cr
e &\rta & g,\quad g \rta - e,\quad\sigma_e \rta \sigma_g,\quad \sigma_g \rta \sigma_e
.\eeqa

Although we have presented these equations as those describing the approximate behaviour 
of a dyonic plasma, they are essentially the same equations that we would have obtained 
for a plasma of separate electric and magnetic charges \cite{ost2}. 
However, the crucial distinction 
between the two cases depends on the nature of the currents $\bJ_e$ and $\bJ_m$. 
In a plasma of separate $e$ and $g$ charges, one finds that these currents are 
linearly independent quantities.
For point particle dyonic plasmas however, the two currents are proportional 
since $e$ and $g$ charges are now localised at the same point, and one finds \cite{ost1}
\beq\label{eq:5}
\bJ_e =\frac{e}{g} \bJ_m. 
\eeq  
Eq.(\ref{eq:5}) has important consequences, in particular it leads to non-trivial DHD
in the so called ideal approximation (where one considers infinite conductivities), which 
we shall discuss shortly. It can be seen that the constraint (\ref{eq:5}) yields the 
following relations 
among $\rho_c,$ $\rho_m,$ $\sigma_e$ and $\sigma_m$
\beq\label{eq:6}
\rho_c=\frac{e}{g}\rho_m, \qquad\qquad\sigma_e =\frac{e^2}{g^2}\sigma_m.
\eeq
Such relations, as well as that of eq.(\ref{eq:5}), are preserved by the duality 
transformations
(\ref{eq:4}). Before discussing the ideal limit of DHD, it is convenient to rewrite 
its equations in a manifestly duality invariant form. Introducing the duality invariant 
quantities
\beqa\label{eq:7}
\cE & = &  e \bE + g\bB\quad \cB=g\bE - e\bB,
\quad \bbJ_e =\frac{\bJ_e}{e},\cr
\brho_c & =& \frac{\rho_c}{e}, \quad \bsig_e =\frac{\sigma_e}{e^2}, 
\eeqa 
the equations of DHD become
\beqa\label{eq:8}
\pat{{\rho}}_{M}+\grad\cdot (\rho_M  \bv  )& = & 0, \quad
\pat { {\brho}}_{c}+\grad\cdot\bbJ_{e}=0, \quad \cr
\cr
c \rho_{M}(\pat{\bv}+\bv\cdot\grad \bv)& = &  
-c  \grad P + c \brho_{c}\cE
- \bbJ_{e}\times\cB, \cr
\cr
\frac{ m_{d}m_{di}}{\rho_M} {\pat {\bbJ}}_{e} & = & 
\frac{m_{d}}{2 \rho_M} \grad P
+ (\cE -\frac{1}{c} \bv\times\cB ) +\frac{m_{d}}{\rho_M c}\bbJ_{e}\times \cB 
-\frac{1}{\bsig_e }\bbJ_{e}, \cr
\cr
\grad\cdot\cB = 0, \quad
\grad\cdot\cE &= &\brho_{c}\lambda, \quad
 \grad\times\cE =\frac{1}{c} \pat{{\cB}}, \quad
 \grad\times\cB  =  -\frac{4\pi}{c} \lambda \bbJ_{e} - \frac{1}{c}\pat {\cE}, 
\eeqa
where $\lambda = e^2 + g^2 $.
The ideal equations of DHD are obtained by considering the limit of infinite 
electric conductivity $\sigma_e \rightarrow\infty$ (and hence also 
$\sigma_m \rightarrow\infty $). The condition that finite electric/magnetic
currents prevail in this limit 
gives rise to the single Ohm's law constraint
\beq\label{eq:9}
\cE  =  \frac{1}{c} \bv \times \cB.
\eeq
Using this constraint, the equations of ideal DHD may be written as 
\beqa\label{eq:10}
\grad \cdot\bv  &=&  0, \qquad\qquad \grad \cdot\cB  = 0,\cr
\grad\times (\bv \times \cB )  &=&  \pat \cB , \qquad
\grad\times ( \pat \bv + \bv \cdot \grad \bv )=  \frac{1}{4 \pi \rho_M 
\lambda}
\times (\cB\cdot \grad \cB).
\eeqa
What is interesting about eqn (\ref{eq:10}) (and their finite conductivity 
 versions) is that modulo factors of $\lambda$ they are equivalent to the 
usual equations of MHD if the duality invariant electric and magnetic fields 
$\cE$ and $\cB$ are associated with the physical ones. The fact that the
 equations of DHD are mappable onto those of MHD makes the task  of 
finding  string-like solutions in the former rather easier, since Olesen 
\cite{olesen}
has already considered magnetic flux tubes in the context of MHD, as well 
as electric flux tubes in the dual  form of MHD. Following \cite{olesen}
 we shall write down an ansatz for a duality invariant, infinitely thin flux tube (which we 
shall call a dyonic flux tube) defined by the field $\cB$.
 Such a string, in the first instance, is described by the embedding 
$Z_i(\sigma,t) \, i = 1,..3 \,$ , where $\sigma$ is a parmeter along the
 flux tube. The field $\cB_i$ is 
\beq\label{eq:11}
{\cB}_i(x,t)={\tilde b}\int 
d\sigma {\partial  Z_i(\sigma,t)\over\partial\sigma}
\delta^3(x-Z(\sigma,t)).
\eeq
where the constant $\tb$ has the interpretation of conserved dyonic flux through an area, 
whose boundary moves with the dyonic fluid velocity. This property of 
the plasma as being \lq\lq frozen to the  field lines" of
$\cB $ is exactly analagous
to the field lines of $B$ in usual plasmas \cite{nicolson}

As usual, in order to obtain (\ref{eq:11}) as a solution to the equations of ideal DHD 
(\ref{eq:10}) we assume that the tube moves with the local dyonic plasma velocity 
${\bf v}(x,t)$, i.e.
\beq\label{eq:12}
v_i(x,t)~\delta^3(x-Z(\sigma,t))
={\partial  Z_i(\sigma,t)\over\pa t}~\delta^3(x-Z(\sigma,t)),
\eeq
with the additional assumption that $\bv$ and $\cB$ are perpendicular
during the flow \cite{olesen}.  It is not necessary to repeat the calculations needed to 
show that  eqs (\ref{eq:11}) and (\ref{eq:12}) do indeed solve the equations of 
DHD (assuming closed dyonic flux tubes), as they follow closely those in
\cite{olesen}. However in passing, we note that a necessary requirement that the duality
 invariant Navier-Stokes  equation in (\ref{eq:10}) be satisfied is that 
\beq\label{eq:13}
P+ \frac{1}{8\pi\lambda}( g^2 E^2 + e^2 B^2 - 2eg\bB\cdot\bE) =
 constant,
\eeq 
which is a generalization of the condition for the existence 
of hydromagnetic waves in MHD, to the case of dyonic plasmas.  
This condition, like that of the flux tube solution (\ref{eq:12}), exhibits 
electric-magnetic duality as defined earlier. Indeed we see that dyonic 
flux tubes are very natural extensions of the purely magnetic or purely 
electric flux tubes considered in \cite{olesen}. If we ignore charge quantization
 (our considerations are purely classical) and consider the limits of either 
$g \rightarrow 0 $ or $e\rightarrow 0$ then dyonic flux tubes as 
solutions to DHD reduce to magnetic tubes in MHD or electric tubes in 
\lq\lq electrohydrodynamics". If we express the physical fields $E _i$and $B_i,$ in 
terms of their duality 
invariant counterparts, we have:
\beqa\label{eq:14}
E_{i} =\mu\Big( \frac{{\epsilon}_{ijk} }{cg}  
v_{j} {\cB}_k +\frac{{\cB}_{i}}{e} \Big), \quad
B_{i} =\mu\Big( \frac{{\epsilon}_{ijk }}{ce}
v_{j} {\cB}_k -
 \frac{{\cB}_{i}}{g}\Big), 
\eeqa
where $\mu={(\frac{\textstyle e}{\textstyle g}+\frac{\textstyle g}{\textstyle e}
{)}^{-1}}.$
Substituting the previous ansatz's for $\cB$ and $\bv$ and using the 
duality invariant Ohm's law constraint (\ref{eq:9}) we obtain
\beqa\label{eq:15}
E_{i} &=&\mu\Big( \frac{{\epsilon}_{ijk} }{cg} {\tilde b} 
\int d\sigma {\partial  Z_j\over\partial t}{\partial  Z_k\over\partial\sigma}
\delta^3(x-Z)+\frac{\tilde b}{e}\int d\sigma {\partial  Z_i\over\partial\sigma}
\delta^3(x-Z).\Big), \cr
\cr
B_{i} &=&\mu\Big( \frac{{\epsilon}_{ijk} }{ce} {\tilde b} 
\int d\sigma {\partial  Z_j\over\partial t}{\partial  Z_k\over\partial\sigma}
\delta^3(x-Z)-\frac{\tilde b}{g}\int d\sigma {\partial  Z_i\over\partial\sigma}
\delta^3(x-Z).\Big),
\eeqa
The above expressions for the physical $\bE$ and $\bB$ fields are
interesting in that they show that each field has a flux tube type
 component linear in the string coordinate $Z_i $ as well as a term
 quadratic in the $Z_i $ which orginates from the motion of the string 
in the dyonic plasma. These latter terms cancel in the expression for $\cB$.
 One consequence of this decomposition into physical $\bE$ and $\bB$ 
fields is that  whilst  the duality invariant flux $\tb$ is time independent as
 discussed above, this is not the case for the separate electric and magnetic
flux inside the flux tube.  If  we denote by $b$ and $b'$ the magnetic and 
electric flux through an area perpendicular to the flux tube, whose
boundary moves with the fluid, then constancy of $\tb$ flux implies
\beq\label{eq:16}
\pat {b'}=  \frac{e}{g} \pat  b,\eeq
so that in principle time varying electric and magnetic fluxes are possible in the 
string as it moves in the dyonic plasama as long as  
variations satisfy (\ref{eq:16}). We shall discuss the typical magnitude  
that such variations might have later, although we can immediately deduce 
that they are proportional to the velocity of the string. This follows
immediately from the fact that in the \lq\lq rest frame"  where $\bv =0,$
$\cE =0$  and we can decompose the dyonic flux tube into 
a separate electric and magnetic tube whose fluxes are related to
$\tb $ via
\beqa\label{eq:17} 
b &= &\int_{A}  \bB \cdot d{\bf A}=-\frac{\tb}{g\mu}, \cr
 \cr
b' &= & \int_{A}  \bE \cdot d{\bf A} =  \frac{\tb}{e\mu}, 
\eeqa
which clearly satisfies the constraint $g b'-e b=\tb$ that follows from
the definition of $\cB$. When the string moves however, there are induced 
electric and magnetic fields, and it is these induced fields that invalidate the above 
picture of dyonic strings being composed of separate electric and
 magnetic strings, in which the separate fluxes are conserved. Rather we 
now have time varying $b'$ and $b$ flux. The question of how large such
 variations  can be is related to another important  issue concerning the
 validity of the  ideal approximation to the equations of DHD (\ref{eq:8}). In
 this limit  \cite{nicolson} not only is $\rho_c \rightarrow 0$ (and hence  $\rho_m 
\rightarrow 0$) but one takes a low frequency approximation
 where for example the term $\pat \cE$ is dropped when compared 
to the term $4\pi\lambda\bJ_e$ in  eqs (\ref{eq:8}).
Using the expression for  $\cE$ obtained in the ideal approximation 
eq (\ref{eq:9}), we can check the validity of dropping such terms.
From the equations of ideal DHD we find
\beqa\label{eq:18}
c  \pat \cE &= &( - \bv \cdot \grad \bv - \frac{1}{\rho_M} 
\grad (P + \frac{ \cB \cdot \cB}{8\pi \lambda} ) 
+ \frac{\cB \cdot \grad \cB }{4 \pi \lambda \rho_M } ) \times \cB\cr 
\cr
&+& \bv \times \grad\times (\bv \times \cB). 
\eeqa
The above equation for $\pat\cE$ can be further simplified using the 
condition that $(P + \frac{ \cB \cdot \cB}{8\pi (e^2 +g^2 )} 
= constant $, (see eq (\ref{eq:13})). Consider now the expression for 
generalized kinetic energy $E_k $ of our dyonic plasma, which is invariant
under electric-magnetic duality
\beq\label{eq:19}
E_k = \int d^3 x ( \frac{\rho_M}{2} \bv \cdot \bv + 
\frac{\cB \cdot \cB }{4 \pi^2 \lambda}).
\eeq
Eq (\ref{eq:19}) tells us that the kinetic energy of the plasma is divided
between velocity and $\cB$  modes of excitation. Indeed if we 
assume we have thermal equilibrium then equipartition arguments
would suggest that such a division is roughly equal, i.e 
\beq\label{eq:20}
\bv  \approx\frac{\cB}{\sqrt{\lambda \rho_M} } 
\eeq
Moreover, barring any accidental cancellations, the Navier -Stokes
equation in (\ref{eq:8}) tells us that
\beq\label{eq:21} 
 | \bJ_e \times  \cB |\approx c \rho_M | \bv \cdot \grad \bv |
\eeq
Hence, we can deduce that 
\beq\label{eq:22}
\frac{ | \pat \, \cE | }{4 \pi \lambda |\bJ_e | } \approx \frac{v^2}{c^2}
\eeq
This makes it clear that the ideal approximation neccesarily implies 
a non-relativistic dyonic fluid, hence non-relativistic strings since 
by assumption the string velocity is identified with that of the 
fluid in which it moves.

Finally, we return to the question of time dependence of the electric and 
magnetic flux $b'(t)$ and $b(t)$ contained in moving dyonic flux tubes, and 
constrained by eq. (\ref{eq:16}). For the time variation of electric
flux one obtains
\beqa\label{eq:23}
\pat b'  &  = & \frac{\mu}{g} \int_{\partial A} ( \bv \times \cB )\cdot
( \frac{\bv}{c} \times d {\bf l} ) - \frac{\mu}{g c} \int d \bA \cdot  \{ ( 
\bv \cdot \grad \bv \cr
\cr
&- & \frac{\cB \cdot \grad \cB }{4 \pi \lambda \rho_M } ) \times \cB - 
\bv \times \grad\times ( \bv \times \cB )  \},
\eeqa  
where in eq (\ref{eq:23}), $d {\bf l}$ is the line element around the 
boundary $\partial A$ that moves with the fluid. In deriving (\ref{eq:23})
we have also used the pressure constraint (\ref{eq:13}). After using 
basic vector identities and the perpendicularity condition 
$\bv \cdot \cB = 0 $ we find
\beqa\label{eq:24}
\pat  b' & =  & \frac{\mu}{g} \int d  \bA \cdot \big[ \{ \frac{\bv}{c}
\times \grad \times ( \bv \times \cB ) - \frac{\bv}{c} \cdot \grad \bv \cr
\cr
& +& \frac{\cB \cdot \grad \cB }{4 \pi \lambda \rho_M c }  \} \times \cB
+ \grad \times ( \frac{v^2 }{c} \cB ) \big]
\eeqa

One may derive a 
similar equation for $\pat b$ directly, or by 
applying  the electric-magnetic duality symmetry to eq (\ref{eq:24}).
Then one finds $\pat b =  g/e \, \pat b'$ in accordance with 
eq (\ref{eq:16}). The crucial observation  to make concerning the 
explicit form of $\pat b$ and $\pat b'$ is that terms which are not small
in the non-relativistic approximation cancel in these expressions. Hence, 
it is clear that the picture of moving dyonic flux tubes as being composed of 
separate electric and magnetic strings is approximately correct in 
non-relativistic dyonic plasmas.

In conclusion, we have considered in this paper dyonic flux tubes in the 
context of generalized 3-dimensional MHD which is invariant under electric-magnetic duality.
They are generalizations of the purely magnetic (electric) strings considered in 
\cite {olesen}, 
and indeed reduce to these in the limit when either  $g$ or $e$ is set to 
zero. We have shown 
that the ideal approximation is valid for  dyonic flux tubes moving in 
non-relativistic dyonic plasmas.

As mentioned in the introduction, a number of obvious questions emerge from our results, 
notably 
what  might be the field theory origin of dyonic flux tubes. Our treatment 
has dealt with dyons in the context of Maxwell theory, and discussed  strings  carrying  
electric and magnetic charge in the usual sense, i.e.  corresponding  
to a single gauge group.

Recently, there has been progress in understanding  strings as solitons of 
superstring theories, in particular as solutions to the low energy supergravity -Yang Mills
system of the compactified theory \cite{dahb}. Since some of these string solutions
are referred to as being \lq\lq dyonic", and although not obviously
connected to the ones we have considered, we mention them for 
purposes of clarification.
The string solutions constructed in \cite{dahb}
are non topological, and carry \lq\lq axionic charge", that is they are described, 
amongst other 
things by a  non trivial three-form field strength $H_{\mu \nu \rho}$
corresponding to the antisymmetric tensor  potential $B_{\mu \nu}$. 
This charge is the analogue 
of electric charge in Maxwell's theory, with invariance under the local $U(1)$ symmetry 
realized as 
$B_{\mu \nu }\rightarrow B_{\mu\nu} + \Lambda_{\mu\nu}$. In 
\cite{schwarz} it was pointed out that the additional 3-form field strength which 
exists in the case of 
type IIb strings, leads to a family of dyonic string solutions whose \lq\lq electric" and 
\lq\lq magnetic" 
charges (related to these two independent 3-form field strengths) 
transform under $SL(2,Z)$.  Here is an example where there are separate 
electric and magnetic  $U(1)$ gauge groups, so as mentioned,
these strings are not directly related to those
described in this paper, but (assuming they can be compactified to d=4) 
they do at  least provide an analogue.

A more likely origin  of dyonic flux tubes, is in a theory
where electric-magnetic duality and its generalizations is realized 
on the same gauge group. This is precisely one of the properties that $S$ duality possesses. 
This symmetry has manifested itself most 
notably in 
rather particular contexts of extended supersymmetry and on dyon-monopole states 
which are BPS saturated \cite{sdual}. These are localized 
solutions rather than the string-like ones we are interested in.
However, just as one may consider magnetic flux tubes as connecting 
magnetic monopole-antimonopole pairs, (possibly leading to the confinement of magnetic 
charge), dyonic flux tubes might naturally connect dyon-antidyon pairs and generate forces 
between them. 
\vskip3cm
{\bf\Large Acknowledgements} 
\vskip0.2in
S. Thomas would like to thank the Royal Society of Great Britain for financial 
support and W. A. Sabra is supported by PPARC.

\end{document}